\documentclass[]{iopart}

\usepackage{amsfonts}
\usepackage{epsf}
\usepackage{graphics}

\begin{document}
\frenchspacing
\title{Applying weighted network measures to microarray distance matrices}
\author{S. E. Ahnert}
\address{Theory of Condensed Matter Group, Cavendish Laboratory, JJ Thomson Avenue, Cambridge CB3 0HE, UK}
\author{D. Garlaschelli}
\address{Dipartimento di Fisica, Universit\'a di Siena,
Via Roma 56, 53100 Siena, Italy}
\author{T. M. A. Fink}
\address{Institut Curie, CNRS UMR 144, 26 rue d'Ulm, 75248 Paris, France}
\author{G. Caldarelli}
\address{INFM-CNR Istituto dei Sistemi Complessi and Dipartimento di Fisica Universit\'a di Roma "La Sapienza" Piazzale Moro 2, 00185 Roma, Italy, and Centro Studi e Museo della Fisica Enrico Fermi, Compendio Viminale, 00185 Roma, Italy}

\begin{abstract}

\noindent
In recent work we presented a new approach to the analysis of weighted networks, by providing a straightforward generalization of any network measure defined on unweighted networks. This approach is based on the translation of a weighted network into an ensemble of edges, and is particularly suited to the analysis of fully connected weighted networks. Here we apply our method to several such networks including distance matrices, and show that the clustering coefficient, constructed by using the ensemble approach, provides meaningful insights into the systems studied. In the particular case of two data sets from microarray experiments the clustering coefficient identifies a number of biologically significant genes, outperforming existing identification approaches.   
\end{abstract}
\maketitle

\noindent
The rise of information technology and the internet, as well as the more recent advent of high-throughput technologies in biology make it easier to obtain large amounts of data on complex networks. Increasingly this also includes data on weighted complex networks, which now appear in many different guises: Transport and traffic \cite{traffic, vespignani}, trade or communication networks, financial networks \cite{kertesz}, and collaboration networks \cite{scn}, to name a few. In biology, genetic regulation and transcription \cite{horvath} and protein interaction \cite{pin} have been studied in this context. However, the extraction of meaningful physical or biological information from these networks is a difficult task. For unweighted complex networks, with binary adjacency matrices, a set of local and global measures on the network has been defined \cite{barabasi}, including the {\em degree} of a node, its {\em average nearest-neighbour degree} \cite{nearest} and its {\em clustering coefficient} \cite{clustering}. Defining these measures for weighted networks is more difficult and has been the subject of recent research \cite{vespignani,horvath,onnela,newman}. A review of definitions of weighted clustering coefficients can be found in \cite{onnela2}. 

In a recent paper \cite{PRE} we introduced a new approach to this problem which allows for a straightforward generalization of any measure defined on an unweighted network to weighted networks. Here we apply the clustering coefficient defined in this way to distance matrices, which are fully connected weighted networks. The distance matrices are generated from microarray expression series, so that closely related series (by some chosen similarity measure) will be separated by a short distance, which in the network picture translates into an edge with a large weight.

The basis of our approach is to find a continuous bijective map $M : \mathbb{R} \rightarrow [0,1]$ from the real numbers to the interval between 0 and 1, which maps the weights $w_{ij} \in \mathbb{R}$ to a quantity $p_{ij} \in [0,1]$. A simple example of such a map is a linear normalization of the weights:
\begin{equation}\label{norm}
p_{ij} = {w_{ij} - {\rm min} (w_{ij}) \over {\rm max}(w_{ij}) - {\rm min}(w_{ij})}
\end{equation}
This simple normalization maps ${\rm min} (w_{ij})$ to zero. While this is often acceptable in the case of a distance matrix, one should make a more sophisticated choice of map if there are many edges with weight ${\rm min} (w_{ij})$. Similarly, if the network has negative weights as well as positive ones, the normalized modulus of the original weights might be a more appropriate choice. A more detailed discussion on the topic of map choice can be found in \cite{PRE}.  

The ideas we introduce in \cite{PRE} are based on an interpretation of the matrix ${\bf P}$ with entries $\{p_{ij}\}$ as a matrix of {\em probabilities}. These probabilities can be interpreted as an {\em ensemble of edges}, or more concisely, an {\em ensemble network}. Thus, just as any binary square matrix can be understood as an unweighted network and any real square matrix corresponds to a weighted network, any square matrix with entries between 0 and 1 corresponds to an ensemble network. If we sample each edge of the ensemble network exactly once, we obtain an unweighted network which we term a {\em realization} of the ensemble network. In particular, $p_{ij}$ is the probability that the edge between nodes $i$ and $j$ exists.
These concepts are valid both for directed networks, with any $p_{ij} \in [0,1]$, and undirected networks, for which $p_{ij} = p_{ji}$, so that the matrix is symmetric. In a real-world weighted network, the original weights can represent almost any physical quantity, such as the strength of a collaboration between two scientists, or the number of passengers traveling between two countries. By mapping these weights to probabilities we rid ourselves of the interpretational burden of these weights, whilst retaining all the topological information they contain. It should be noted that in many cases the interpretation of weights as probabilities also makes intuitive physical sense. Whenever the weights in a network represent a magnitude of flow, this can be interpreted directly in terms of the probability that a transfer occurs during a given unit of time. Examples include traffic and transport networks as well as communication networks, where we have units (passengers, money, signals) which form an edge, through their transfer, with a probability proportional to the flow rate. 

All measures on unweighted networks can be written as functions of the entries $a_{ij}$ of an adjacency matrix ${\bf A}$. In fact, generally they can be written as a polynomial of these entries, or a simple ratio of such polynomials. Note that, for an unweighted network, $a_{ij} = a_{ij}^m$ for all positive integers $m > 0$, so that these polynomials are of first order only. Consider a general first-order polynomial, which can be written fully expanded as:
\[
f({\bf A}) = \sum_{q = 0}^{2^{N^2}} C_q \prod_{j,k = 0}^{N} a_{jk}^{b(q)_{jk}} 
\]
where $N$ is the number of nodes, the $C_q$ are real coefficients and the $b(q)_{jk}$ are a set of boolean matrices specifying which adjacency matrix entries appear in each term of the polynomial.  
The probability $P_q$ that $\prod_{j,k = 0}^{N} a_{jk}^{b(q)_{jk}} = 1$ in a given realization ${\bf A}$ is simply $P_q = \prod_{j,k = 0}^{N} p_{jk}^{b(q)_{jk}}$. Thus, due to the linearity of the polynomial, the average $\bar{f}({\bf P})$ of $f$ over the ensemble network realizations is:
\begin{equation}\label{poly}
\bar{f}({\bf P}) = \sum_{q = 0}^{2^{N^2}} C_q \prod_{j,k = 0}^{N} p_{jk}^{b(q)_{jk}} = f({\bf P})
\end{equation}
This means that the value of a polynomial function $f$ of the entries of an unweighted network ${\bf A}$, averaged over the realizations of a given ensemble network ${\bf P}$ is equal to the value of the polynomial of the ensemble network adjacency matrix itself. 

The degree $k_i$ of a given node $i$ in an unweighted network with adjacency matrix elements $a_{ij}$ is the number of its neighbours, and is written as $k_i = \sum_{j} a_{ij}$. 
In a weighted network with elements $w_{ij}$ the corresponding quantity has been termed the {\em strength} of the node $i$, denoted as $s_i$, which consists of the sum of the weights: $s_i = \sum_{j} w_{ij}$.
In an ensemble network, the corresponding sum over the edges attached to a particular node gives the {\em average degree} of node $i$ {\em across realizations}, denoted as $\bar{k_i}$ and given by $\bar{k_i} = \sum_{j} p_{ij}$.
 
It is important to note that while the strength of a node in a weighted network may have meaning in the context of the network, $\bar{k_i}$ has a universal meaning, regardless of the original meaning of the weights. 

As a more complex example, consider the {\em clustering coefficient} of a node $i$, which has been defined \cite{clustering} as:
\begin{equation}\label{uclusteringeq}
c_i = {\sum_{j,k} a_{ij} a_{jk} a_{ik} \over k(k-1)/2} = {\sum_{j,k} a_{ij} a_{jk} a_{ik} \over \sum_{j,k} a_{ij} a_{ik}} 
\end{equation}
where $k \neq j \neq i \neq k$ in the sums. This corresponds to the number of triangles in the network which include node $i$, divided by the number of pairs of bonds including $i$, which represent potential triangles. Using the ensemble approach with its normalized weights this generalizes straightforwardly to:
\begin{equation}\label{clusteringeq}
c^e_i = {\sum_{j,k} p_{ij} p_{jk} p_{ik} \over \sum_{j,k} p_{ij} p_{ik}} 
\end{equation}
which can be read as the average number of triangles divided by the average number of bond pairs. In modified form, this clustering coefficient has appeared in the very recent literature \cite{horvath} but without connection to a general approach to the construction of weighted network measures based on a general mapping from weights to probabilities. 
Note that $c^e_i$ is {\em not} the average of $c_i$ over the ensemble. For a detailed discussion of this subtlety, see \cite{PRE}.

\begin{figure}
\begin{tabular}{ccc}
\hspace{-0.6cm}\scalebox{0.2}[0.2]{\rotatebox{-90}{\epsfbox{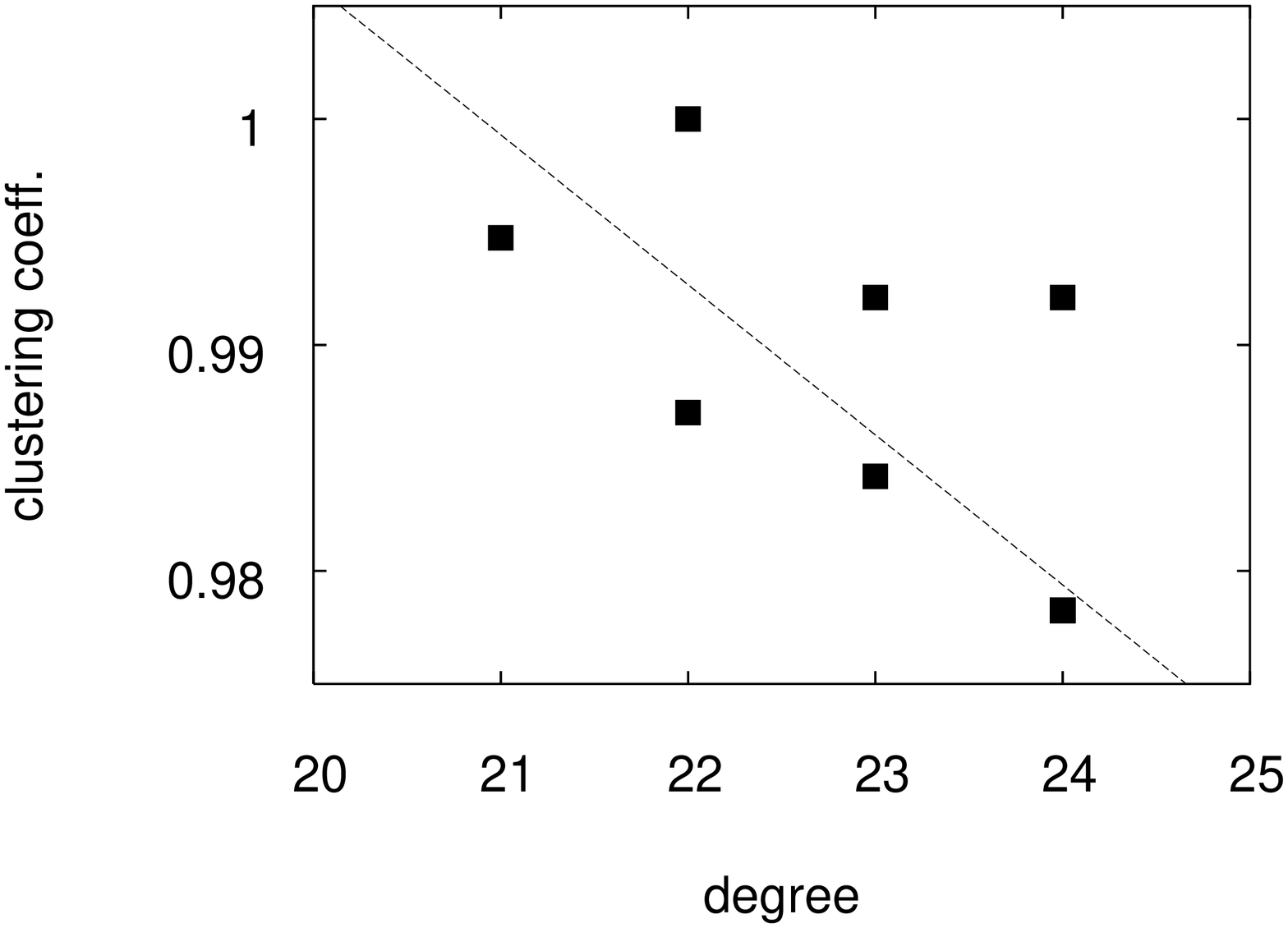}}}&
\hspace{-0.8cm}\scalebox{0.22}[0.2]{\rotatebox{-90}{\epsfbox{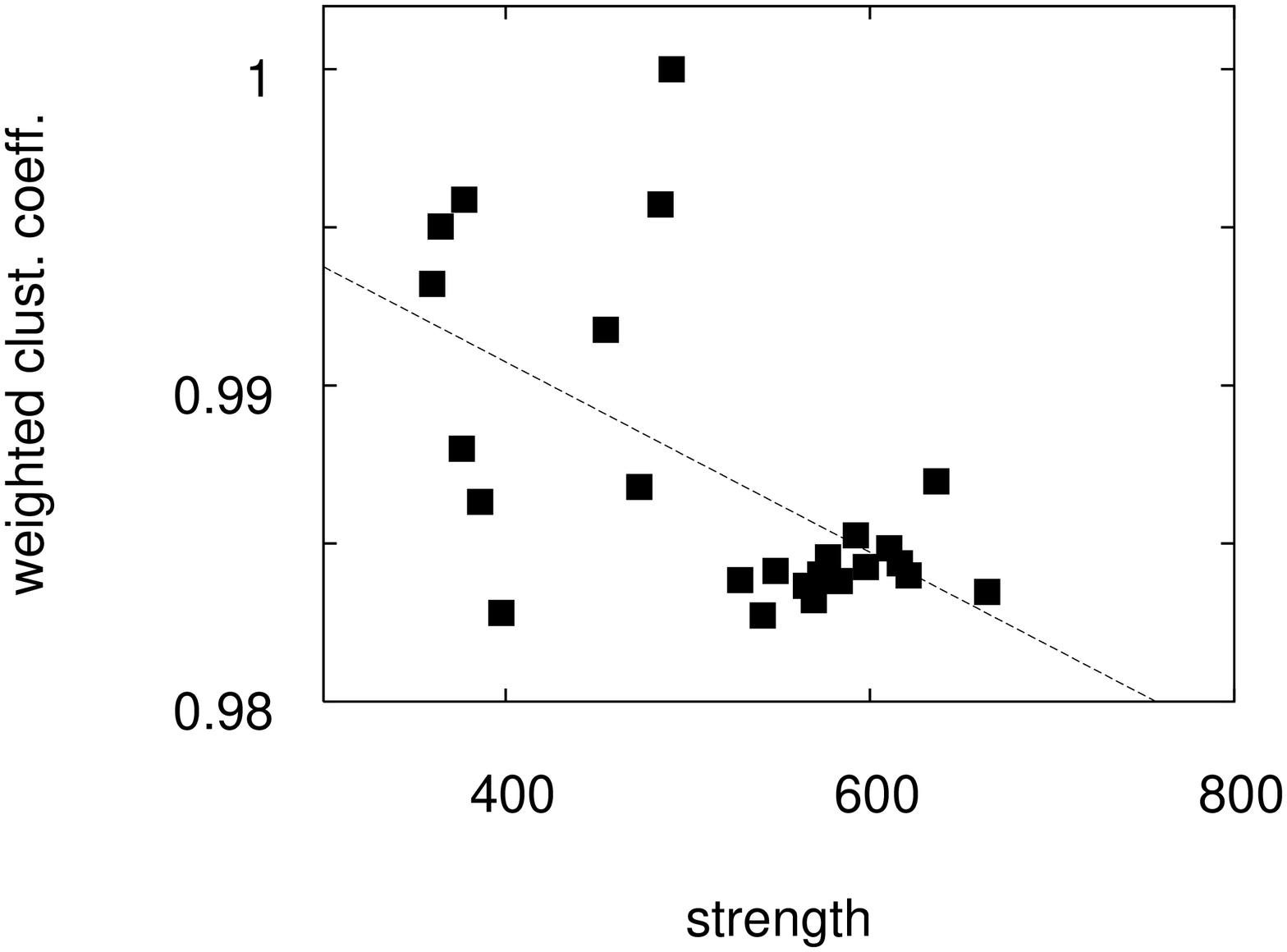}}}&
\hspace{-0.3cm}\scalebox{0.2}[0.2]{\rotatebox{-90}{\epsfbox{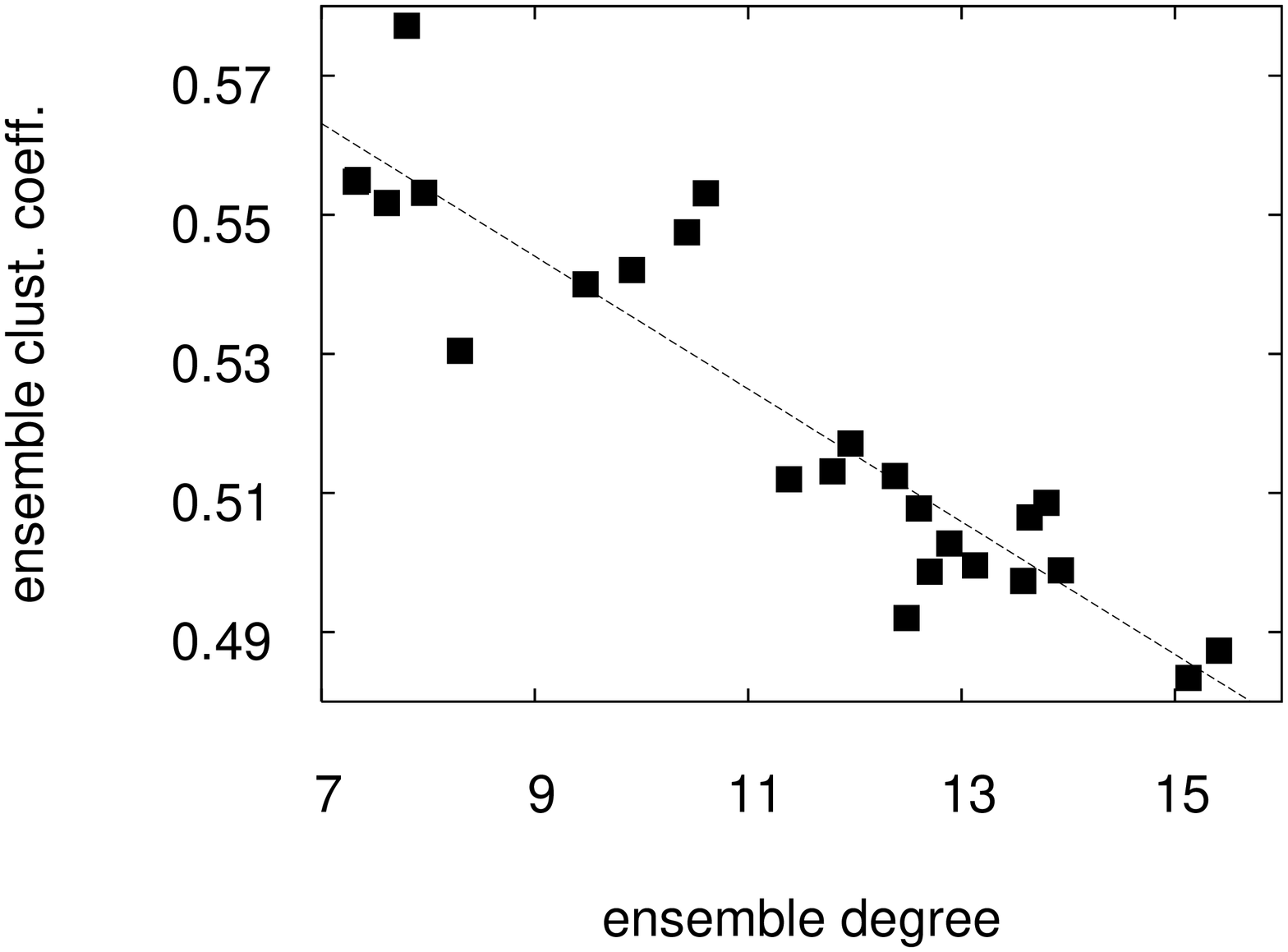}}}\cr
\end{tabular}
\caption{Example of the advantages of the ensemble clustering coefficient, as shown in our earlier work \cite{PRE}: The network of air travel passengers within the 25 member states of the EU\cite{eurostat} is almost fully connected. LEFT: Unweighted clustering coefficient versus degree.All 25 data points are projected onto 7 locations, as a result of the information loss due to discarding the weights, and because the network is almost fully connected. CENTER: Clustering coefficient as proposed in the literature \cite{vespignani} versus strength. This ``mixed'' clustering coefficient is a function of unweighted and weighted quantities. No clear relationship is evident, again because the network is almost fully connected. RIGHT: Ensemble clustering coefficient versus ensemble degree. Unlike the other two approaches, those derived using the ensemble quantities exhibit a clear negative linear relationship. The lines are lines of best fit. Note that the absolute scale of the ensemble clustering coefficient $c^e_i$ depends on the choice of the map $M$ from weights to probabilities, which makes the relative values of $c^e_i$ more important than the absolute ones.}\label{eu}
\end{figure} 

All measures constructed with the ensemble approach are only functions of the normalized weights $p_{ij}$, not of the elements of an unweighted adjacency matrix $a_{ij}$ or of the degree $k$. This distinguishes the ensemble measures from measures proposed for weighted networks in the literature, such as the weighted clustering coefficient $c_i^w$:
\begin{equation}\label{vesp}
c_i^w = {1 \over s_i (k_i - 1)} \sum_{j,k} {(w_{ij} + w_{ik}) \over 2} a_{ij} a_{ik} a_{jk}
\end{equation}
and the weighted average nearest-neighbour degree $k_{nn,i}^w$:
\begin{equation}\label{vesp2}
k_{nn,i}^w = {1 \over s_i} \sum_{j=1}^N a_{ij} w_{ij} k_j
\end{equation}

Both are defined in \cite{vespignani}, and eq. (\ref{vesp}) is the most frequently cited definition of a weighted clustering coefficient in the literature. Due to their construction, these measures cannot be used for the analysis of fully connected weighted networks, as $k_{nn,i}^w = 1$ and $c_i^w = 1$ for all nodes $i$ in such networks. Fully connected weighted networks form an important class of complex networks, for example in the form of the (virtually fully-connected) EU air travel network which we analyze in \cite{PRE} (see Fig. \ref{eu}). Furthermore, {\em any} matrix of similarities or distances between a number of objects - such as for instance microarray data series in biological experiments - can be treated as a fully connected weighted network, and thus can be analyzed using the ensemble approach, but not with approaches such as eq. (\ref{vesp}) and (\ref{vesp2}), which are ``mixed'' in the sense that they make use of both the unweighted and weighted adjacency matrix entries. 

Note that the absolute values of the ensemble clustering coefficient have limited meaning, as they are dependent on the map $M$. It is their relative values which carry the information, and these are largely independent of the choice of map $M$, as long as it is bijective. 

\begin{figure}
\begin{tabular}{cc}
\hspace{-0.8cm}\scalebox{0.35}[0.35]{\rotatebox{-90}{\epsfbox{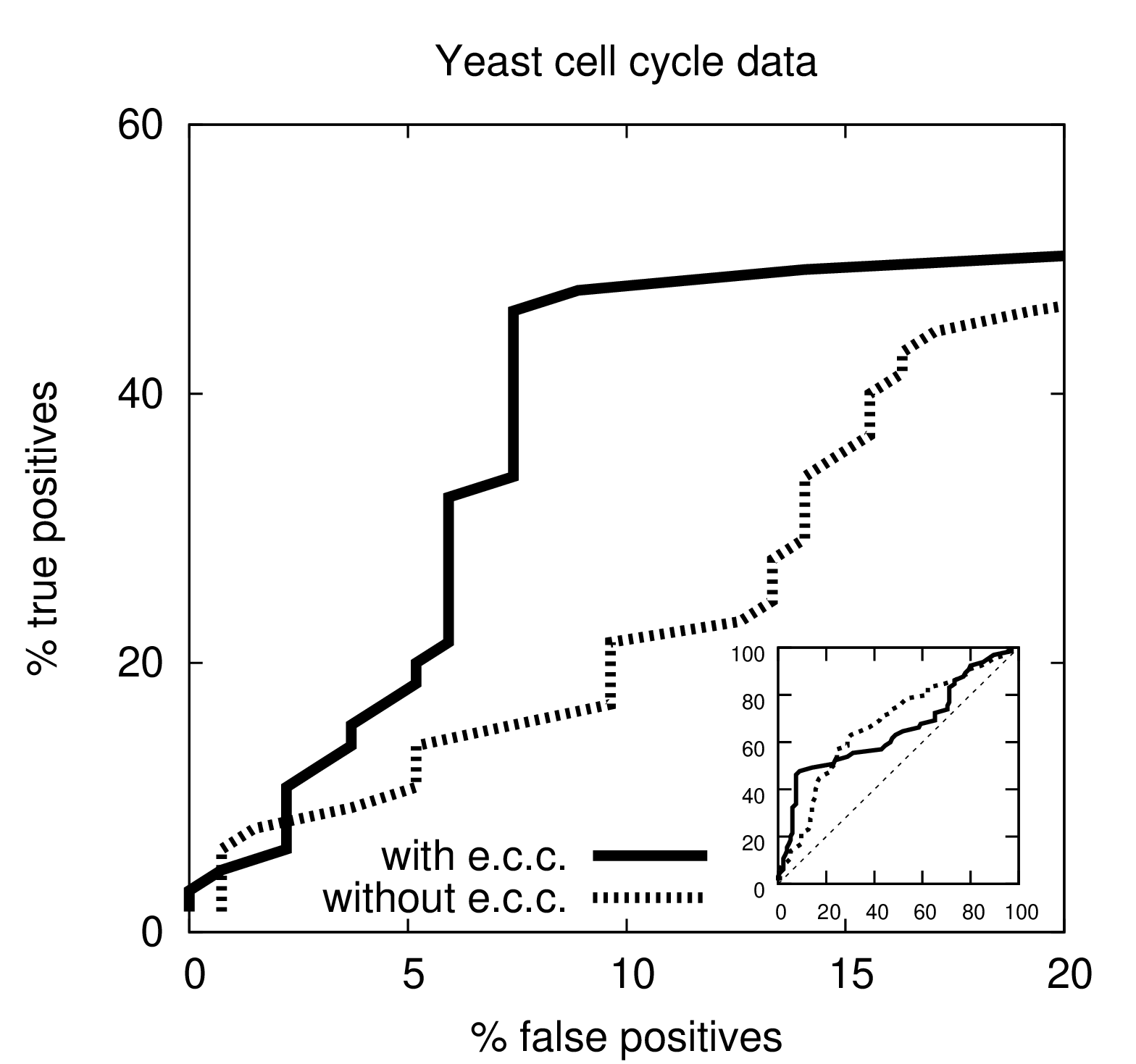}}}&
\hspace{-0.3cm}\scalebox{0.35}[0.35]{\rotatebox{-90}{\epsfbox{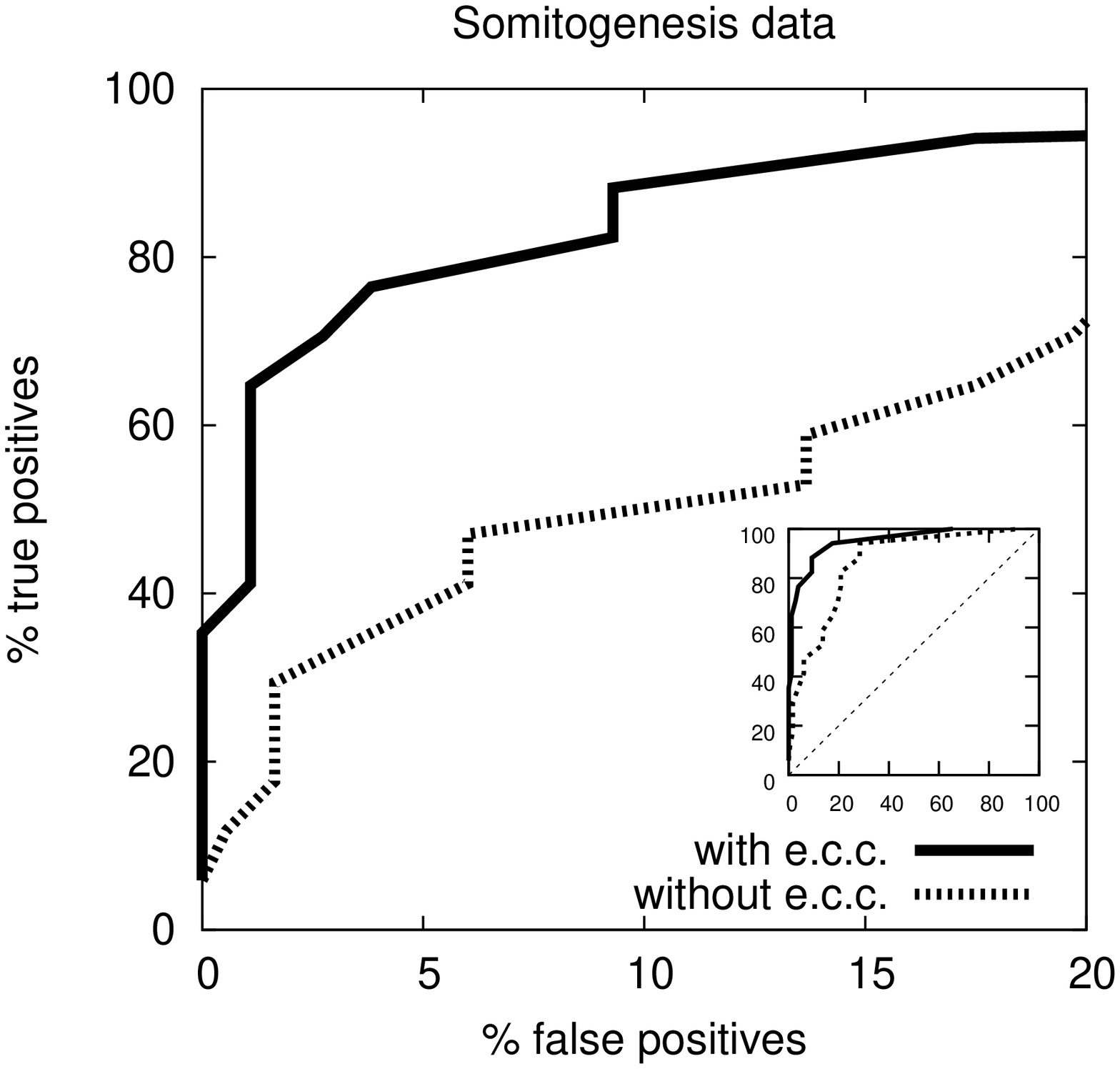}}}\cr
\end{tabular}
\caption{Receiver-operating characteristic (ROC) diagrams for the yeast cell cycle (LEFT) and somitogenesis (RIGHT) datasets, showing the positions of known biologically significant genes in a ranking of 200 genes in the rankings generated (a) using the ensemble clustering coefficient (solid) and (b) using the original pattern-finding approach (dotted) which was used to select the 200 genes in the first place. In both cases the ensemble clustering coefficient moves biologically significant genes to the top of the ranking.}\label{roc}
\end{figure} 

Microarrays are one of the most successful high-throughput technologies in biology, providing a snapshot of gene expression levels for all of the thousands of genes in the genome of a given organism simultaneously. A microarray consists of a large number of microscopic spots on a slide (typically made of glass or silicon), which each contain copies of a different short DNA sequence (or {\em oligonucleotide}) unique to a particular gene. Furthermore, the sequence copies in each spot are attached to a flourescent marker. If a given gene is expressed in the tissue sample to be examined, many copies of this gene will be present in the form of messenger RNA (mRNA), which in turn will bind to the sequences on the microarray, causing flourescence of the spot. The flourescence of the array of spots is captured by a camera and then read out using a computer. 

A series of microarray measurements gives an expression profile for each gene over space or time, telling us where and when a given gene is 'switched on'. These sets of data series are subjected to detailed analysis, and distance matrices between these series, (often calculated using Pearson correlation) typically form an integral part of such an analysis. 

Here we calculate the ensemble clustering coefficient for distance matrices derived from two entirely different microarray data sets. The first data set consists of microarray data from an experiment studying the formation of vertebra (somitogenesis) in mice \cite{Dequeant06}, from which a list of 200 genes was compiled using an existing pattern detection approach \cite{bioinfo}. This approach is designed to detect biologically significant genes by finding expression profiles which deviate from randomness. The second data set is the well-known dataset of yeast cell cycle microarray experiments in yeast \cite{Spellman}. Here too the 200 strongest patterns were selected using the same approach. 

It should be noted that microarray datasets are notoriously noisy and pre-filtering of data based on purely mathematical measures is essential and in fact present in almost any microarray study. Our selection method based on pattern detection is mathematically rigorous and makes no prior assumptions about the nature of the pattern. 

In each of the two datasets the 200 genes are ranked by the amount of pattern they contain (and thus by their supposed biological significance). Yet the fully connected weighted network which corresponds to a distance matrix between these 200 genes contains none of this information. Therefore, when we calculate the ensemble clustering coefficient for a distance matrix of 200 genes, we can use the pattern-detection approach as a benchmark comparison for the performance of the clustering coefficient in finding biologically significant genes.  

For both the mouse somitogenesis and yeast cell cycle datasets we compare our predictions to lists of known biologically significant genes. In the case of mouse somitogenesis these are 17 genes associated with the Wnt and Notch pathways, listed in \cite{Dequeant06}, and in the case of yeast cell cycle there are 65 genes which can be found in two lists of experimentally verified yeast cell cycle genes \cite{Spellmanknown,Simon}. 
 
The distance measure chosen to generate the distance matrix is the algorithmic compression of one expression series due to another \cite{bioinfo}. As can be seen in Fig. \ref{roc}, the ranking generated by using the clustering coefficient clearly outperforms the pattern-ranking for both datasets. In the case of the mouse somitogenesis dataset, 11 (64\%) of the 17 genes known to play a role in somitogenesis are located in the top 13 places (top 6\%) of the ranking. Similarly, in the yeast cell cycle dataset, 31 (48\%) of 65 known genes occupy places in the top 43 (top 21\%). Compared to this, the conventional pattern-finding approach fares less well, with 6 (35\%) in the top 13 (somitogenesis) and 23 (35\%) in the top 43 (yeast). The conclusion is that in both datasets the ensemble clustering coefficient appears to move biologically significant genes to the top of the ranking.

By transforming a weighted network into an ensemble network, any of the numerous measures which have been defined for unweighted networks can be straightforwardly generalized to weighted networks. As we have shown in this paper, our approach is particularly suited for the analysis of distance matrices. We demonstrate this by calculating the ensemble clustering coefficient for the distance matrices between microarray data series which successfully identifies many known biologically significant genes. These results are an indication that the application of complex networks methods to the rather separate field of distance matrix analysis is likely to yield valuable insights.

\end{document}